\documentclass{camera}
\usepackage{axodraw}

\begin{document}
\def\ga{\gamma}
\newcommand{\be}[1]{\begin{equation}\label{#1}}
\newcommand{\ee}{\end{equation}}
\renewcommand{\to}{\rightarrow}
\newcommand{\beqn}[1]{\begin{eqnarray}\label{#1}}
\newcommand{\eea}{\end{eqnarray}}
\def\bY{{\bf Y}}
\def\tl{{\tilde{L}}}
\def\bM{{\bf M}}
\def\bm{{\bf m}}
\def\al{\alpha}
\def\la{\lambda}
\def\bU{{\bf U}}
\def\lsim{\raise0.3ex\hbox{$\;<$\kern-0.75em\raise-1.1ex
\hbox{$\sim\;$}}}
\def\gsim{\raise0.3ex\hbox{$\;>$\kern-0.75em\raise-1.1ex
\hbox{$\sim\;$}}}
%
%
\title{LEPTON FLAVOUR VIOLATION IN \\
\vspace{0.1cm}
SUPERSYMMETRIC MODELS}

%
\author{ANNA ROSSI}

%

\organization{
Dipartimento di Fisica `G.~Galilei',  
Universit\`a di Padova \and \\
\vspace{0.07cm}
INFN, Sezione di Padova,
Via Marzolo 8, I-35131 Padua, Italy \\
\vspace{0.1cm}
{\it Talk given at IFAE 2003, April 23-26, 2003 - Lecce, Italy. }}
\vglue -0.5cm
\maketitle

%
Nowadays the only evidence of lepton flavour violation (LFV) is 
provided by neutrino oscillations, which are the 
result of non-vanishing neutrino masses and mixing angles. 
It would be very important to observe 
other LFV phenomena. The Standard Model, however, does not offer 
promising scenarios since, for example, the LFV prototype decays, 
{\it i.e.} $\mu \to \ga e, \tau\to \mu \ga~ {\rm etc.}$ are strongly 
suppressed by the smallness of the neutrino masses. 
The Minimal Supersymmetric extension of the  Standard Model (MSSM) 
is, on the other hand, a natural framework where several such processes 
could be sizeable through the one-loop exchange of superpartners, 
if the masses of the latter are not too heavy and do not conserve 
flavour \cite{EN}. 
In this contribution
 we shall present two recent developments on 
LFV in supersymmetric models. In the first part (${\bf 1.}$) we will discuss 
the supersymmetric seesaw mechanism induced by the exchange of heavy 
$SU(2)_W$-triplet states ($T$), in alternative to `right-handed' 
neutrinos ($N$), to generate neutrino masses.
We will show that LFV can be radiatively induced in the slepton masses and 
that this scenario is potentially more predictive than the standard 
$N$-seesaw \cite{AR}.  
In the second part (${\bf 2.}$) we shall focus on a new LFV signal, namely the 
decays of the neutral Higgs bosons, $(h^0,H^0,A^0)\to \mu \tau$ \cite{BR}.

\begin{figure}[htb]

\begin{center}
\begin{picture}(340,80)(0,0)
\Text(20,40)[]{(a)}
\Text(55,40)[]{$\bY_N$}
\Text(125,40)[]{$\bY_N$}
\Text(40,10)[bl]{$L$}
\Text(40,75)[tl]{$H_2$}
\Text(90,30)[]{$N$}
\Text(90,50)[]{$\bM_N$}
\Text(140,10)[br]{$L$}
\Text(140,75)[tr]{$H_2$}
\ArrowLine(50,20)(70,40)
\ArrowLine(50,60)(70,40)
\ArrowLine(90,40)(70,40)
\ArrowLine(90,40)(110,40)
\ArrowLine(130,20)(110,40)
\ArrowLine(130,60)(110,40)
\ArrowLine(270,0)(290,20)
\ArrowLine(310,0)(290,20)
\ArrowLine(290,40)(290,20)
\ArrowLine(270,80)(290,60)
\ArrowLine(310,80)(290,60)
\ArrowLine(290,40)(290,60)
\Text(240,40)[]{(b)}
\Text(260,0)[b]{$L$}
\Text(320,0)[b]{$L$}
\Text(260,80)[t]{$H_2$}
\Text(320,80)[t]{$H_2$}
\Text(300,30)[]{$T$}
\Text(280,25)[]{$\bY_T$}
\Text(280,42)[]{$M_T$}
\Text(280,60)[]{$\la_2$}
\Text(300,50)[]{$\overline{T}$}
\end{picture}
\end{center}
\vskip -0.5 cm
\caption{\small Seesaw mechanisms for the neutrino mass generation:
(a) heavy `right-handed' singlet exchange;  (b) heavy triplet exchange.}
\label{f1}
\end{figure}
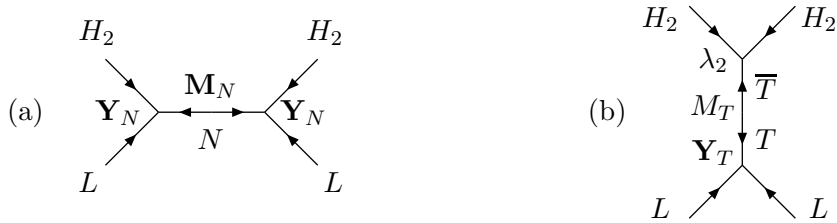

{\large \bf 1.} {\underline {\it LFV without right-handed neutrinos.}}   
Perhaps the best way to catch the difference between the 
$N$-induced seesaw and the $T$-induced one is to look at  
Fig.~1 where we have drawn the diagrams responsible for the 
neutrino mass generation in the former (a) and latter case (b). 
The main difference we are now interested in  regards the flavour structure 
involved in the two scenarios. 
In the standard seesaw we have two flavour sources as 
shown in Fig.~1: ${\bf Y}_N$,  a  $3 \times 3$ arbitrary  
Yukawa matrix,  and ${\bf M}_N$,  a $3 \times 3$ symmetric mass 
matrix. The low-energy neutrino masses are given by $
{\bf m}^{ij} _\nu  =v_2^2~ 
\bY^{T ik}_N{\bf M}^{-1 k l}_N\bY^{l j }_N$ ($\langle H_2\rangle=v_2= 
v\sin\beta, ~ v= 174~{\rm GeV}$). 
A simple counting of the parameters demonstrates that  
the low-energy ones, described by ${\bf m}_\nu$, 
which amount to 6 real parameters + 3 phases, are less than 
the number of the independent parameters in $\bY_N$ 
and $\bM_N$, which instead are 12 real parameters + 6 phases. 
We cannot infer from the low-energy data the more fundamental 
quantities, $\bY_N$  and  $\bM_N$. 
In the $T$-seesaw we have  a single flavour source, namely the 
symmetric matrix $\bY_T$. 
In this case the latter is directly related to the neutrino masses, 
 ${\bf m}^{ij}_\nu = 
\frac{v_2^2 \lambda_2}{M_T} \bY^{i j}_T$. As a result 
the independent $\bY_T$ parameters, 6 real parameters + 3 phases, 
are just matched by the low-energy physical parameters. 
Now let us come to the main point. 
It is well-known that $\bY_N$ 
can induce lepton-flavour violating  entries in the 
mass matrices ${\bf m}_{\tilde{L}}$ 
of the left-handed sleptons through radiative 
corrections \cite{old}, 
even in the minimal SUSY scenario 
with universal soft-breaking terms at the GUT scale $M_G$. 
In the $N$-seesaw scenario the size of this LFV cannot be 
 unambiguously predicted in a bottom-up approach making use of the low-energy 
data (neutrino masses and mixing angles). On the other hand,  in the 
 $T$-seesaw scenario
the LFV entries can be directly connected  
 to the effective neutrino mass matrices \cite{AR}:
\be{Tsusy1}
  ( \bm^{2 }_{\tilde{L}})_{ij} \propto  
 m^2_0 (\bY^{\dagger}_T \bY_T)_{ij} {\log}\frac{M_G}{M_T}
\sim m^2_0 
\left(\frac{M_T}{\la_2 v^2_2}\right)^2 (\bm^{\dagger}_\nu \bm_\nu )_{ij}
{\log}\frac{M_G}{M_T}
 , ~~~~
i\neq j . 
\ee
Therefore, the relative size  of LFV 
in the 1-2, 2-3, 1-3  sectors  can be approximately predicted in terms 
of only the low-energy masses and neutrino mixing angles, {\it i.e.}:
\be{predi}
 \frac{ (\bm^{2 }_{\tilde{L}})_{\tau \mu}}
  {(\bm^{2 }_{\tilde{L}})_{\mu e} } \approx \left(
\frac{m_3}{m_2}\right)^2 \frac{\sin 2\theta_{23}} {\sin 2\theta_{12}
\cos\theta_{23}} \sim 80 ,~~~~ \frac{ (\bm^{2 }_{\tilde{L}})_{\tau e}}
  {(\bm^{2 }_{\tilde{L}})_{\mu e} }\approx 
\tan\theta_{23} \sim 1 , 
\ee
where  the numerical estimates use the values of  
$\theta_{12}, \theta_{23}$, 
$m_2, m_3$ indicated by the solar and atmospheric solutions \cite{nus}.
These estimates can be translated into 
a prediction for the ratios of the decay rates 
of $\tau \to \mu \ga, \tau \to e \ga$ and $\mu\to e \ga$: 
\beqn{ratepredi1}
\frac{BR(\tau \to \mu\ga)}{BR(\mu \to e\ga)} & \approx & 
\left(\frac{(\bm^2_{\tl})_{ \tau \mu}}{(\bm^2_{\tl})_{\mu e}}\right)^2 
\frac{BR(\tau \to \mu \nu_\tau \bar{\nu}_\mu)}
{BR(\mu \to e \nu_\mu \bar{\nu}_e)} \sim 10^3  \nonumber \\
\frac{BR(\tau \to e \ga)}{BR(\mu \to e\ga)} &\approx &
\left(\frac{(\bm^2_{\tl})_{ \tau e}}{(\bm^2_{\tl})_{  \mu e}}\right)^2 
\frac{BR(\tau \to e \nu_\tau \bar{\nu}_e)}
{BR(\mu \to e \nu_\mu \bar{\nu}_e)} \sim 10^{-1}  . 
\eea
More detailed numerical computations have confirmed the above estimates 
\cite{AR} demonstrating remarkably that these ratios depend only on 
low-energy $\nu$ parameters and not on the details of the models\footnote{
The $T$-seesaw scenario can be 
embedded in a $SU(5)$ scenario  in which the 
states $T$ fill the 15-representation together 
with other coloured partners. In such a case quark-flavour violation 
can be linked to LFV   \cite{AR}.}, such as 
soft-breaking parameters or the triplet mass $M_T$. 

{\bf 2.} {\underline {\it LFV Higgs couplings.}} In the MSSM, 
by integrating out the sleptons, gauginos and 
Higgsinos, LFV 
(dimension-four) interactions arise:
\be{LFV}
{\cal L}_{FV} = - \frac{Y_\tau}{\sqrt2 \cos\beta} 
( \Delta_L ~\tau^c \mu +\Delta_R~ \mu^c \tau )~ 
[ h^0 \cos(\beta -\alpha)
-H^0 \sin(\beta -\alpha) - {\rm i} A^0] 
+ {\rm h.c.}   
\ee
where   $\Delta_L, \Delta_R$  
are  dimensionless functions of the MSSM mass parameters, 
$\alpha$ is the  mixing angle in the CP-even Higgs sector.
The branching ratios can be expressed as ($\Phi^0= h^0,H^0, A^0$):
\be{rphi}
{ BR(\Phi^0\to \mu^+\tau^-)}
=  \tan^2\beta~ (|\Delta_L|^2 + |\Delta_R|^2) 
 ~ C_\Phi~ { BR(\Phi^0\to \tau^+\tau^-)}  \, ,
\ee
where $C_A=1$ and  $C_{h, H}$ depend on $\beta$ and  $\al$ \cite{BR}. 
Interesting effects can be obtained for large LFV violation 
[{\it e.g.} $(\bm^2_{\tl})_{\mu \tau}\! \sim \! (\bm^2_{\tl})_{\mu \mu} \! 
\sim \! (\bm^2_{\tl})_{\tau \tau}]$ and large 
$\tan\beta$ values, say $\sim \!50$, which can yield 
$|50\Delta|^2\sim 10^{-3}$ \cite{BR}.  
The best phenomenological prospects are found for the decays of the 
non-standard Higgs bosons, corresponding to 
$H^0, A^0$ ($h^0, A^0$) 
for $m_A \gsim m_\star$ ($m_A \lsim m_\star$), where  
$m_\star \sim 110 - 130~{\rm GeV}$. For these bosons 
we may obtain 	$BR(\Phi^0\to \mu^+\tau^-) \sim 10^{-4}$, which 
corresponds to 
about $(10^4,10^3, 2\cdot 10^2)$ events at LHC for 
  $m_A \sim(100,200,300)~{\rm GeV}$ and 
 an integrated luminosity of $100 ~{\rm fb}^{-1}$.
We recall that  the main production mechanisms of those bosons at LHC 
are  bottom-loop mediated gluon fusion 
and associated production with $b \bar{b}$.

Another important issue regards  possible correlations 
between  the decays  $\Phi^0\to \mu\tau$ and other LFV processes. 
For large $\tan\beta$ and large LFV also the decay rate 
for $\tau \to \mu \gamma$ 
is enhanced and could exceed the experimental limit. 
Since the rate of $\tau \to \mu \gamma$ decreases 
as the superparticle masses increase, whereas the rate of 
$\Phi^0\to \mu \tau$ does not, it is sufficient to 
 raise such masses  towards the TeV region to fulfill the bound.
The decays  $\Phi^0\to \mu\tau$ are also correlated to 
the decay $\tau \to 3\mu$, which is due to both dipole contributions 
via $\gamma$ exchange  
and LFV scalar operators (\ref{LFV})  via Higgs exchange \cite{BK}.
As for the Higgs-mediated contribution, we obtain the 
following estimate \cite{BR}:  
\be{tau3}
BR(\tau\to 3\mu)_{\Phi^*} \sim 10^{-7} 
\left(
\frac{\tan\beta}{50}\right)^6  
\left(\frac{100 ~{\rm GeV}}{m_{A}}\right)^4 ~
\left( \frac{|50 \Delta_L|^2 +|50 \Delta_R|^2}{10^{-3}} \right)  ~.
\ee
Therefore, this contribution 
can exceed the dipole induced one, 
$BR(\tau\to 3\mu)_{\ga^*} \sim 2.3\times 10^{-3} ~ BR( \tau\to \mu \ga) 
\lsim 7 \times 10^{-10}$,  
and be not far from 
the present bound, $BR(\tau \to 3 \mu) < 3.8 \times 10^{-7}$.
Notice that the  parameter region in which this occurs  
is also the most favorable one for the observation of 
the $\Phi^0 \to \mu \tau$ decays, so an interesting correlation 
emerges.

\vspace{0.1cm}

\noindent
{\bf Acknowledgments} ~{\small  This work  was   partially supported 
by MIUR and 
the European Union under the contracts 
HPRN-CT-2000-00148 (Across the Energy Frontier) and
HPRN-CT-2000-00149 (Collider Physics).}



\vskip -2cm

\begin{thebibliography}{99}
\bibitem{EN}
J.~Ellis and D.~V.~Nanopoulos,
Phys.\ Lett.\  {\bf B110} (1982) 44.

\bibitem{AR}
A.~Rossi, Phys.\ Rev.\ D {\bf 66} (2002) 075003.


\bibitem{BR}
A.~Brignole and A.~Rossi,
Phys.\ Lett.\ B {\bf 566} (2003) 217.

\bibitem{old} 
F.~Borzumati and A.~Masiero, Phys. \ Rev. \ Lett. {\bf 57}, 961 (1986); 
F.~Gabbiani and A.~Masiero, Nucl. \ Phys. B {\bf 322} (1989) 235, and 
more recently: 
J.~Hisano and K.~Tobe,
Phys.\ Lett.\ B {\bf 510} (2001) 197; 
J.~A.~Casas and A.~Ibarra,
Nucl.\ Phys.\ B {\bf 618} (2001) 171; 
A.~Masiero, S.~K.~Vempati and O.~Vives,
Nucl.\ Phys.\ B {\bf 649} (2003) 189.


\bibitem{nus}
S.~Fukuda {\it et al.}  [Super-Kamiokande Collaboration],
Phys.\ Rev.\ Lett.\  {\bf 85} (2000) 3999;
G.~L.~Fogli, E.~Lisi, A.~Marrone and A.~Palazzo,
arXiv:hep-ph/0309100;
M.~Maltoni, T.~Schwetz, M.~A.~Tortola and J.~W.~F.~Valle,
arXiv:hep-ph/0309130;
P.~C.~de Holanda and A.~Y.~Smirnov,
arXiv:hep-ph/0309299. 

\bibitem{BK} 
K.~S.~Babu and C.~Kolda,
Phys.\ Rev.\ Lett.\  {\bf 89} (2002) 241802; 
A.~Dedes, J.~R.~Ellis and M.~Raidal, 
Phys.\ Lett.\ B {\bf 549} (2002) 159.

\end{thebibliography}
\end{document}